\documentstyle[prb,twocolumn,aps]{revtex}
\input{epsf}
\newcommand{\bm}[1]{\mbox{\boldmath$#1$}}
\begin{document} \draft 
\twocolumn[
\title{Phase separation as an instability of the Tomonaga-Luttinger liquid}
\author{Masaaki Nakamura\cite{email1} and Kiyohide Nomura\cite{email2}}
\address{Department of Physics, Tokyo Institute of Technology,
Oh-Okayama, Meguro-ku, Tokyo 152, Japan}
\date{\today}\maketitle 
\begin{abstract}
\widetext\leftskip=0.10753\textwidth \rightskip\leftskip
Asymptotic behavior of the Tomonaga-Luttinger liquid
in the vicinity of the phase-separated region is investigated
in the one-dimensional $t$-$J$ model,
to study the universal property of the $c=1$ conformal field theory
with U(1) symmetry near the $K\rightarrow\infty$ instability.
On the analogy of the spinless fermion,
we discuss that the compressibility behaves as $\kappa\propto (J_c-J)^{-1}$,
and that the Drude weight is constant and
changes to zero discontinuously at the phase boundary.
This speculation is confirmed by analyzing the finite size effect
from the result of the exact diagonalization.
\end{abstract}
] \narrowtext
\section{Introduction}

Phase separation in one-dimensional (1D) electron systems
has attracted attention of condensed matter physicists,
since it is observed that the superconducting correlation
is enhanced near the phase-separated region.
Many electron systems show phase
separation\cite{Ogata-L-S-A,Hellberg,Penc&Mila,Sano&Ono},
of which the $t$-$J$ model is the simplest.
The Hamiltonian of the 1D $t$-$J$ model is written as
\begin{eqnarray}
{\cal H}&=&-t\sum_{i\sigma}(c^{\dag}_{i\sigma} c_{i+1\sigma}
                   +c^{\dag}_{i+1\sigma} c_{i\sigma})\nonumber\\
 &&+J\sum_{i}(\bm{S}_i\cdot\bm{S}_{i+1}-n_i n_{i+1}/4),\label{eqn:t-J}
\end{eqnarray} 
in the subspace without double occupancy.
In general, unless some instability occurs,
1D electron systems belong to
Tomonaga-Luttinger liquids (TLL)\cite{Solyom,haldane-tl},
which is characterized by gapless charge and spin excitations
and power-law decay of correlation functions.
The charge part of the excitation is described by the $c=1$
conformal field theory (CFT) with U(1) symmetry,
while the spin part is described by the $c=1$ CFT with SU(2) symmetry.
The dominant correlations are determined by a single exponent $K_{\rho}$.
In the phase diagram of the 1D $t$-$J$ model obtained
by Ogata {\it et al.}\cite{Ogata-L-S-A},
the charge or the spin density fluctuations are dominant ($K_{\rho}<1$)
for the small $J/t$ region, while $J/t$ is increased,
the superconducting fluctuations become dominant ($K_{\rho}>1$).
Finally, the compressibility diverges at $J_c/t=2.5$-$3.5$
($K_{\rho}\rightarrow\infty$)
and the system goes into the phase-separated state for $J>J_c$.

As a typical instability of the $c=1$ U(1) CFT,
Berezinskii-Kosterlitz-Thouless (BKT)
transition\cite{B-K-T} has been well known.
In electron systems, the Mott transition belongs to this type\cite{Giamarchi}.
However, the asymptotic behavior of the $c=1$ U(1) CFT with
$K\rightarrow\infty$ has not been fully investigated.
In 1D spin systems, this type of instability corresponds to the transition
between the massless $XY$ phase and the ferromagnetic phase.
In this article, we argue the asymptotic behavior in the vicinity of
the instability $K\rightarrow\infty$,
relating it with the 1D spinless fermion system.
Then we apply our speculation to the 1D $t$-$J$ model
by analyzing the finite size effect
from the result of the exact diagonalization.

This paper is organized as follows.
In section \ref{sec:CFT}, we discuss an instability of the $c=1$ CFT
with U(1) symmetry, and review several physical examples.
In section \ref{sec:num}, we numerically calculate physical quantities
in the 1D $t$-$J$ model to examine the argument.
Finally, a summary is given in section \ref{sec:sum}.

\section{Phase separation and the $c=1$ U(1) CFT}\label{sec:CFT}

The $c=1$ CFT with U(1) symmetry is described by the Gaussian model
\begin{equation}
 {\cal H}=\frac{1}{2\pi}\int dx \left[vK(\pi\Pi)^2+\frac{v}{K}
 \left(\frac{\partial\phi}{\partial x}\right)^2\right],
\end{equation}
where $\Pi$ is the momentum density conjugate to $\phi$,
$[\phi(x),\Pi(x')]=i\delta(x-x')$, $K$ is the Gaussian coupling,
and $v$ is the sound velocity.
Its dual field $\theta$ is defined as $\partial_x \theta(x)= \pi\Pi(x)$.
We make the identification
$\phi \equiv \phi + 2 \pi/\sqrt{2},
\theta \equiv \theta + 2 \pi/\sqrt{2}$,
which means the U(1) symmetry for the $\theta$ field.
The scaling dimensions for the operators
\begin{equation}
        O_{n,m} = \exp (i n \sqrt{2} \phi) \exp (i m \sqrt{2} \theta)
\end{equation}
are
\begin{equation}
        x_{n,m} = \frac{1}{2} \left( n^2 K + \frac{m^2}{K} \right).
        \label{eq:gauss-critical}
\end{equation}
Therefore, the scaling dimension $x_{0,m}$ decreases to
$0$ for $K \rightarrow \infty$, 
which implies phase separation in electron systems,
or a ferromagnetic long-range order in spin systems.

For a finite $L$ size system with periodic boundary conditions, 
the excitation energies are related to 
the scaling dimensions \cite{Cardy}
\begin{equation}
        \Delta E_{n,m} (L) = \frac{2\pi v}{L} x_{n,m},
\end{equation}
and the correction to the ground state energy is described 
by the conformal charge $c$ 
\cite{blote-cardy}
\begin{equation}
        E_0 (L) = e_0 L - \frac{\pi v}{6 L} c,
\end{equation}
where $e_0$ is the bulk energy density.

\subsection{Spinless fermion systems}

As the simplest physical model,
we consider the spinless fermion with nearest neighbor interactions
($t$-$V$ model),
\begin{equation}
 {\cal H}=-t\sum_{i}(c^{\dag}_i c_{i+1}+c^{\dag}_{i+1} c_i)
 -V\sum_i n_i n_{i+1}.\label{eqn:t-V}
\end{equation}
For simplicity, we replace the parameters as $\Delta\equiv V/2|t|$.
Then the metallic phase, which is described by the $c=1$ U(1) CFT,
becomes unstable and phase separation ($K\rightarrow\infty$)\cite{Haldane-sf}
occurs for the region of $\Delta \geq 1$ at any fillings.

The quantum numbers in (\ref{eq:gauss-critical})
correspond as $(n,m)=(\Delta D,\Delta N)$
\cite{Bogoliubov,Pokrovskii-T,Woynarovich-E},
where $\Delta D$ denotes the number of particles moved
from the left Fermi point to the right one,
and $\Delta N$ is the change of the total number of particles.
The selection rule under periodic boundary conditions is
\begin{equation}
\Delta D=\frac{\Delta N}{2}\ (\mbox{mod}\,1).
\end{equation}

Since the maximum values of the $\Delta N$ and $\Delta D$
are of the order of the system size,
and the excitation energies $\Delta E_{\Delta D,0}=\pi\Delta D^{\,2}vK/L$
are limited by the band width ($\propto L$),
so that $vK$ is expected to be constant near the critical point
($K\rightarrow\infty$).

It is thought that the transition between the metallic state
and the phase-separated state is the first order,
because the ground state of the phase-separated state
(the state where the band is filled up or empty,
i.e., $\Delta N=\pm L/2$ at half-filling),
and the metallic state ($\Delta N=0$) is thermodynamically different.
Therefore, the excited energy corresponding to the phase-separated state
from the metallic ground state should be
\begin{equation}
\Delta E_{0,\pm L/2}=\frac{\pi v}{L}\frac{(L/2)^2}{K}\propto (1-\Delta)L,
\label{eq:1st-order}
\end{equation}
which means $v/K\propto 1-\Delta$.

In fact, using the Jordan-Wigner transformation,
the Hamiltonian (\ref{eqn:t-V}) at half-filling
is mapped onto the $S=1/2$ $XXZ$ model with the anisotropy
$\Delta$\cite{Luther}.
From the Bethe ansatz result \cite{cloizaux}, we obtain
\begin{equation}
        1/K = \frac{1}{\pi} \arccos\Delta
        \approx \frac{1}{\pi} \sqrt{2(1-\Delta)},
\end{equation}
and
\begin{equation}
        v = \frac{\pi}{2}\frac{\sin (\arccos\Delta)}{\pi- \arccos\Delta}
        \approx \frac{1}{2} \sqrt{2(1-\Delta)}.
\end{equation}
Therefore, near $\Delta=1$, the sound velocity behaves 
$v \propto 1/K$, as expected.
Note that the phase-separated state
has the excited energy near $\Delta=1$
\begin{equation}
        \Delta E_{0,\pm L/2}
        = \frac{L}{4}(1-\Delta),
\end{equation}
which is consistent with (\ref{eq:1st-order}).
Thus, near $\Delta=1$, the asymptotic behaviors are
$vK\sim\mbox{Const.}$,$v/K\propto 1-\Delta$, as expected.

\subsection{Electron systems}

In the case of 1D electron systems, generally,
the charge and the spin degrees of freedoms are separated,
and the charge part is described by the $c=1$ U(1) CFT,
while the spin part is described by the SU(2) CFT
\cite{Woynarovich,Frahm-K,Kawakami&Yang90b},
\begin{equation}
 \Delta E = \frac{2\pi v_c}{L}x_c +  \frac{2\pi v_s}{L}x_s.
\end{equation}
Here the scaling dimensions are given by
\begin{mathletters}
\begin{eqnarray}
x_c&=&\frac{1}{2}\left[
		  \frac{(\Delta N_c/2)^2}{K_{\rho}}
		  +K_{\rho}(2\Delta D_c+\Delta D_s)^2\right],\label{eqn:xc}\\
x_s&=&\frac{1}{2}\left(\Delta N_s-\frac{\Delta N_c}{2}\right)^2
    +\frac{1}{2}\Delta D_s^{\ 2},
\end{eqnarray}
\end{mathletters}
where
the subscripts $c,s$ denote the charge and the spin
degrees of freedoms respectively,
and $\Delta N_s$ means the change of the number of the down spins.
These quantum numbers are restricted by the selection rule
under periodic boundary conditions\cite{Woynarovich}:
\begin{mathletters}
\begin{eqnarray}
\Delta D_c&=&\frac{\Delta N_c+\Delta N_s}{2}\ (\mbox{mod}\ 1),\label{eqn:dDa}\\
\Delta D_s&=&\frac{\Delta N_c}{2}\hspace{1.3cm} (\mbox{mod}\ 1).\label{eqn:dDb}
\end{eqnarray}
\end{mathletters}

Since both the spinless fermion (the $S=1/2$ $XXZ$ spin chain)
and the charge part of the electron system,
for example the 1D $t$-$J$ model, belong to the same universality class,
we expect that the asymptotic behavior in the limit of $K\rightarrow\infty$
is the same for both cases.
Therefore, the charge velocity and the exponent for the charge part
are expected to behave as $v_c K_{\rho} \sim \mbox{Const.}$,
$v_c / K_{\rho}\propto J_c-J$ respectively.

Next, we relate $v_c$ and $K_\rho$ to the compressibility
and the Drude weight.
From the universal relations of TLL\cite{Schulz,Kawakami&Yang90a},
the compressibility is given by
\begin{equation}
 \frac{1}{n^2\kappa}=\frac{\pi}{2}\frac{v_c}{K_{\rho}},\label{eqn:kappa}
\end{equation}
so that the compressibility is expected to behave like
$\kappa\propto (J_c-J)^{-1}$.
On the other hand, the Drude weight is given by the relation
\begin{equation}
D=\frac{K_{\rho}v_c}{\pi},\label{eqn:drude}
\end{equation}
therefore, it is constant (at least finite) at the phase boundary.
In the phase-separated state,
the system is separated into an electron rich and a hole rich regions,
and the former is regarded as an antiferromagnetic Heisenberg chain
with open boundary conditions,
so that the dc conductivity in this non-uniform state is expected to be $0$.
This indicates that the Drude weight changes discontinuously
at the phase boundary.

\subsection{Spin systems}

For a model of spin systems, we treat the $XXZ$ spin chain
\begin{equation}
{\cal H}=-\sum_{j}(S^x_j S^x_{j+1}+S^y_j S^y_{j}+\Delta S^z_j S^z_{j+1}).
 \label{bond-alterXXZ}
\end{equation}
This model has a massless $XY$ phase close to the $\Delta < 1$, 
and a ferromagnetic phase for $\Delta \ge 1$.

As mentioned in the previous subsection,
the Hamiltonian (\ref{bond-alterXXZ}) for $S=1/2$
is obtained by the Jordan-Wigner transformation
of the $t$-$V$ model (\ref{eqn:t-V}).
However, the boundary condition depends on the number of particles
in the spinless fermion system: $S^{\pm}_{L+1}=e^{\pm i\pi N}S^{\pm}_{1}$,
which changes the selection rule to the bosonic one,
\begin{equation}
\Delta N=\mbox{integer},\ \Delta D=\mbox{integer}.
\end{equation}
Changing the selection rule,
our argument for the spinless fermion also holds in this case.
The fact that $v\propto 1/K$ explains the change of the dispersion curve from 
the type $\omega \approx v |q|$ in the $XY$ region 
to the $\omega \propto q^2$ on the SU(2) ferromagnet, 
and that the ground state energy of the ferromagnet does not
depend on the system size.  

For the general spin $S$ case, in the $XY$ phase 
the fully ferromagnetic states ($S^z_T = \pm SL$) have the excited energy 
\begin{equation}
        \Delta E_{0,\pm SL} = S^2 L (1-\Delta)[1 +O(1-\Delta,1/L)],
        \label{eq:1st-o2}
\end{equation}
(see Appendix \ref{apx:A}).
Comparing this with the Gaussian model arguments
(\ref{eq:gauss-critical}), we obtain 
\begin{equation}
        \Delta E_{0,\pm m} = \frac{m^2}{L} (1-\Delta) 
        = \frac{\pi v}{L}\frac{m^2}{K}.
        \label{eq:1st-o3}
\end{equation}
Therefore, assuming $v \propto 1/K$ near $\Delta=1$, 
we obtain $v, 1/K \propto \sqrt{1-\Delta}$.  
And the energy gap between the fully ferromagnetic state and 
the one-spin flip state ($S^z_T = \pm (SL-1)$) is
\begin{eqnarray}
        \lefteqn{\Delta E_{0,\pm SL}-\Delta E_{0,\pm (SL-1)}}\nonumber\\
        && = \frac{2SL-1}{L}(1-\Delta) \approx 2S (1-\Delta),
\end{eqnarray}	
which is consistent with the simple spin wave calculation.

For the ferromagnetic region ($\Delta\geq 1$),
there is a large degeneracy $O(L)$ in the ground state
under the special boundary condition,
which reflects the invariance of the
Hamiltonian under the quantum group\cite{Alcaraz-S-W}.
Physically, this degeneracy is related
with the translational invariance of the domain wall.

\section{Numerical calculation}\label{sec:num}

To examine the above prediction, we diagonalize the Hamiltonian
of the $t$-$J$ model (\ref{eqn:t-J})
using the Lanczos method for 8,12,16,20 sites clusters
at quarter filling.
In the finite size calculation,
the compressibility $\kappa$ is given by
\begin{equation}
 \kappa=
\frac{L}{N^2}\left(\frac{E_0(L;N+2)+E_0(L;N-2)-2E_0(L;N)}{4}\right)^{-1},
\label{eqn:kappa_num}
\end{equation}
where $E_0(L;N)$ is the ground state energy of a system
with size $L$ and $N$ electrons ($n\equiv N/L$).
We choose periodic boundary conditions for $N=4m+2$ ($m$: integer) electrons,
and antiperiodic boundary conditions for $N=4m$ electrons.
The reason of this choice is as follows.
When the number of electrons are changed by 2 ($\Delta N_c=2$)
keeping $S_T^z=0$, the number of up spin changes by 1 ($\Delta N_s=1$).
Then, from the selection rule (\ref{eqn:dDa}),
the possible value of $\Delta D_c$ shifts by $1/2$.
The change of the boundary condition cancels this phase shift,
and it makes the ground state
always singlet with zero momentum\cite{Ogata&Shiba,Ogata-L-S-A}.

The Drude weight \cite{Kohn,Stephan-H} is given with the relation
\begin{equation}
 D=\frac{L}{2}\left.\frac{\partial^2E_0(\Phi)}{\partial\Phi^2}\right|_{\Phi=0},
\end{equation}
where the flux-dependent ground state energy $E_0(\Phi)$ is calculated
by modifying the hopping part of (\ref{eqn:t-J}):
\begin{equation}
 -t\sum_{j,\sigma}e^{i\Phi/L}c^{\dag}_{j\sigma}c_{j+1\sigma}+\mbox{H.c.}.
\end{equation}
By fitting the energy difference $E_0(\Phi)-E_0(0)$
as a function $A\Phi^2+B\Phi^4$,
we find that $A\gg |B|>0$.
Therefore, the Drude weight and $\Phi^2$
is well approximated by the linear relation,
and we can calculate the Drude weight by the difference of energies
(e.g., at $\Phi=0$ and $0.1\pi$).
This is a natural consequence in the TLL,
because $\Delta D_c$ is modified as $\Delta D_c+\Phi/2\pi$ for $\Phi\neq 0$,
so that (\ref{eqn:xc}) gives only an $O(\Phi^2)$ dependence
if the ground state is singlet.
The Drude weight is known exactly at $J/t=0$ as $D=\sin(\pi n)/\pi$,
and at $J/t=2$ from the Bethe ansatz\cite{Kawakami&Yang91}.
We have checked the value at these points and confirmed
the validity of our calculation.

As we expected, the inverse compressibility $\kappa^{-1}$
vanishes linearly toward the phase boundary in the TLL region
(see FIG.\ref{fig:comp1}).
The finite size effect is very small,
but it has an $O(L^{-2})$ size dependence.
This correction is explained by the irrelevant fields $(x=4)$
\cite{Cardy86,Reinicke}
\begin{equation}
 L_{-2}\bar{L}_{-2}{\bf 1},\ \ 
 \left\{(L_{-2})^2+(\bar{L}_{-2})^2\right\}{\bf 1}.
\end{equation}
Although the compressibility should be non-negative
in the thermodynamic limit ($L\rightarrow\infty$),
there exists a region with $\kappa^{-1}<0$.
The reason for this phenomenon is considered as follows:
we have done the computation under the micro canonical ensemble,
while the theory of TLL is constructed in the grand canonical ensemble.
The negative inverse compressibility is expected to approach $0$
for larger systems (see FIG.\ref{fig:ksizep}).

FIG.\ref{fig:Drude1} shows that the Drude weight decreases to zero
as $J/t$ is increased,
but the value remains finite at the phase boundary $J=J_c$
determined by $\kappa^{-1}=0$.
There is also an $O(L^{-2})$ size dependence for $J<J_c$.
However, for $J>J_c$, the slopes of the Drude weight become steeper
as the system size increases, and it seems to have
a discontinuity in the thermodynamic limit.
This is consistent with our discussion that
the Drude weight remains finite at the phase-separation boundary.

In order to check the consistency of the relation for the TLL,
we compare the charge velocities obtained by the two independent methods
(see FIG.\ref{fig:velsc1});
one is obtained from the low-energy spectrum as
\begin{equation}
 v_c=\frac{E_1(L;N;S=0)-E_0(L;N)}{2\pi/L},
 \label{eqn:vc}
\end{equation}
where $E_1(L;N;S=0)$ is the first singlet excited state for
the wave number $2\pi/L$, and the other is given by the relation derived
from (\ref{eqn:kappa}) and (\ref{eqn:drude}),
\begin{equation}
 v_c=\sqrt{\frac{2D}{n^2\kappa}}\label{eqn:vc2}.
\end{equation}
We have seen that the Drude weight and the compressibility
have the little size dependence for $J<J_c$.
It is natural that the size dependence of the charge velocity
obtained by (\ref{eqn:vc}) is larger than that of (\ref{eqn:vc2}),
since the former is
given by the difference of the energies for discrete wave numbers,
but the latter is given by the differentiation of the energy
by the continuous quantity $\Phi$,
so that we used finite size data (at $L=16$) for (\ref{eqn:vc2}).
We find that the charge velocities calculated in the two ways
are consistent in the region $J/t=2.0$-$3.0$,
extrapolating (\ref{eqn:vc}) as
$v_c(L) = v_c(\infty) + A/L^2 + B/L^4$.

Finally, to see the degeneracy in the phase-separated region,
we do the Legendre transformation $f\equiv e-\mu n$,
where $e \equiv E_0 / L$.
The chemical potential $\mu$ is defined by
\begin{equation}
 \mu(L;N)=\frac{E_0(L;N+2)-E_0(L;N-2)}{4}\label{chem_pot}.
\end{equation}
The $f$ versus $J/t$ at quarter-filling is shown
in FIG.\ref{fig:first}. As the system size $L$ is increased,
there appear two regimes with different slopes,
and the value $f$ approaches $0$ in the phase-separated region
(see also FIG.\ref{fig:fsizep}).
This is the evidence that the transition is the first order.
In fact, the phase-separation boundary $J_c$ can also be defined at the
point of $f = 0$ which is equivalent to
the Maxwell construction\cite{Emery-K-L}.
In finite systems, the Maxwell construction tends to estimate the 
phase boundary at smaller $J$ than the one determined by $\kappa^{-1}=0$
\cite{Ogata-L-S-A,Hellberg-Manousakis}.

FIG.\ref{fig:grand_ene} shows that $f$ is almost constant against
the change of the electron density in the phase-separated region.
This result suggests that
the phase-separated state approximately has degeneracy $O(L)$,
which corresponds to the large degeneracy $O(L)$ in the
ferromagnetic region ($\Delta\geq 1$) of the spin $S$ $XXZ$ model.

\section{Summary and discussions}\label{sec:sum}

In this paper, we have studied the instability of the $c=1$ CFT 
for the U(1) symmetry case.
For the spinless fermion (or the $S=1/2$ $XXZ$ spin chain),
considering that the excitation
energies are limited by the band-width, 
we showed that the sound velocity is inverse to the
Gaussian coupling $v \propto 1/K$.
In addition, since the transition
from the metallic ($XY$) phase to the phase-separated (ferromagnetic) state
is the first order, the spin wave velocity and the Gaussian coupling behave
as $vK\sim\mbox{Const.}, v/K \propto 1-\Delta$.

About the 1D $t$-$J$ model, since the low-energy behavior of the charge
part is described by the $c=1$ U(1) CFT, and the critical exponent 
$K_\rho$ diverges near the phase separation, 
we expected that the compressibility is proportional to $(J_c-J)^{-1}$,
and the Drude weight is constant.
The obtained numerical results supports this expectation.

These asymptotic behaviors of the TLL have been observed in many cases;
the linear behavior of the inverse compressibility was also observed
by Troyer {\it et al.}\cite{Troyer}
in the 1D extended $t$-$J$ model with the (next-)nearest-neighbor repulsion.
The discontinuity of the Drude weight (kinetic energy) was found by
Sandwik and Sudb{\o} in the case of the 1D two band Hubbard model,
with increasing the nearest-neighbor repulsion\cite{Sandvik}.
Our argument gives a unified interpretation of these behaviors.
In particular, the latter phase-separated state is complicated,
and similar to the one which was found by Sano and \={O}no\cite{Sano&Ono}.
Although these phase separations are different from the 1D $t$-$J$ model,
the universality class is expected to be the same, so that
our argument can be applied to these cases,
and the discontinuity of the Drude weight and other asymptotic behavior
can be explained in the same way.

Historically, the similar asymptotic behavior in $K\rightarrow\infty$
was discussed by the g-ology\cite{Solyom}.
However, in the exactly solvable electron systems such as the simple
Hubbard model or the supersymmetric $t$-$J$ model ($J/t=2$),
there is no phase separation, therefore such a possibility has not been
considered seriously.

Although our speculation is successful in explanation
of the behavior of the TLL with $K\rightarrow\infty$,
the higher corrections to $vK$, $v/K$ near the critical point
are quite ambiguous.
There are complicated effects come from the surface energy.
It will be useful to investigate this behavior
by analyzing the $t$-$V$ model
where the exact solution is available.

\section{Acknowledgments}
The authors are grateful to M. Ogata, H. Shiba, and A. Kitazawa
for valuable suggestions and discussions.
They also thank C. S. Hellberg for communications about
the Maxwell construction.
This work is partially supported by Grant-in-Aid for
Scientific Research (C) No. 08640479 from the Ministry of Education,
Science and Culture, Japan.
The computation in this work was done
using the facilities of the Supercomputer Center,
Institute for Solid State Physics, University of Tokyo.
\appendix
 \section{}\label{apx:A}

Here we show that the relation (\ref{eq:1st-order}) applies not only to the
integrable case $S=1/2$ but also to the non-integrable case, $S$ arbitrary.  

Close to $\Delta=1$, we can estimate the ground state energy 
by the perturbation.  
We treat the SU(2) ferromagnetic term 
\begin{equation}
        {\cal H}_0=-\sum \mbox{\boldmath$S$}_j \cdot
        \mbox{\boldmath$S$}_{j+1},
\end{equation}
as a free part, and the anisotropic part
\begin{equation}
        {\cal H}_1=-(\Delta-1) \sum S^z_j S^z_{j+1},
\end{equation}
as a perturbation.  

The energy of the fully ferromagnetic state $S^z_T= \pm SL $ is exactly 
\begin{equation}
        E_{ferro} = - \Delta  S^2 L.
\end{equation}
For $\Delta <1$, the ground state has the quantum number $S^z_T=0,q=0$.  
At $\Delta=1$, the ground state wave function in the $S^z_T=0,q=0$ 
space is derived from the fully ferromagnetic state
\begin{equation}
        | \phi \rangle = (S^-_T)^{SL}| S^z_T=SL\rangle,
\end{equation}
where $S^\pm_T \equiv \sum S^{\pm}_j$.  
The zero-th order energy is given by 
\begin{equation}
        {\cal H}_0 |\phi \rangle = -S^2 L | \phi \rangle.
\end{equation}
To calculate the first order perturbation, 
we have to evaluate 
$\langle\phi| S^z_j S^z_{j+1} |\phi\rangle$.  
Since $| \phi \rangle$ is invariant under the permutation of 
the lattice sites $\{ j \}$, we obtain
\begin{equation}
        \langle \phi | S^z_i S^z_j | \phi\rangle = \mbox{Const}. 
        \qquad
        \mbox{for any $i \neq j$},
\end{equation}
therefore, 
\begin{eqnarray}
\lefteqn{\langle \phi | ( \sum S^z_i )^2 | \phi\rangle = 0}
        \nonumber       \\
        &=& \sum_{i\neq j} \langle \phi | S^z_i S^z_j | \phi\rangle 
        + \sum_i \langle \phi | (S^z_i )^2 | \phi\rangle
        \nonumber       \\
        &=& L(L-1)  \langle \phi | S^z_j S^z_{j+1} | \phi\rangle 
        + L \langle \phi | (S^z_i)^2 | \phi\rangle.  
\end{eqnarray}
Considering 
$
\langle \phi |(S^z_i)^2|\phi\rangle /\langle \phi|\phi \rangle \propto
S^2
$,
we obtain 
$
\langle\phi| S^z_j S^z_{j+1} |\phi\rangle
/\langle \phi|\phi \rangle \propto -S^2/L
$.  
Then, the first order perturbation is 
\begin{eqnarray}
        E_1 &=& \langle \phi | {\cal H}_1 | 
        \phi\rangle /\langle \phi | \phi \rangle        \nonumber       \\
        &=& -(\Delta-1)
        \sum \langle \phi| S^z_j S^z_{j+1} | \phi \rangle
        /\langle \phi| \phi \rangle
        \nonumber       \\
        &\propto& -(\Delta -1) S^2.
\end{eqnarray}
Therefore, the energy gap between the singlet ground state and the
fully ferromagnetic state is
\begin{equation}
        \Delta E_{0,\pm SL} = S^2 L (1-\Delta) 
        [1 +  O(1-\Delta,1/L)].
\end{equation}

\begin{figure}
\epsfxsize=3.3in \epsfbox{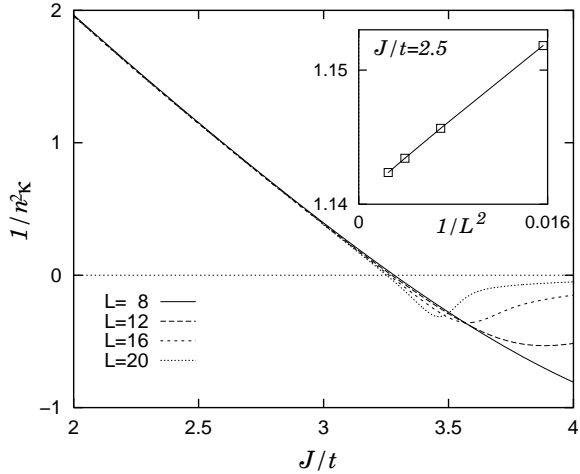}
\caption{Compressibility $\kappa$ as a function of $J/t$
for different system sizes at $n=1/2$.
The size dependence at $J/t=2.5$ is shown in the inset.}\label{fig:comp1}
\end{figure}
\begin{figure}
\epsfxsize=3.3in \epsfbox{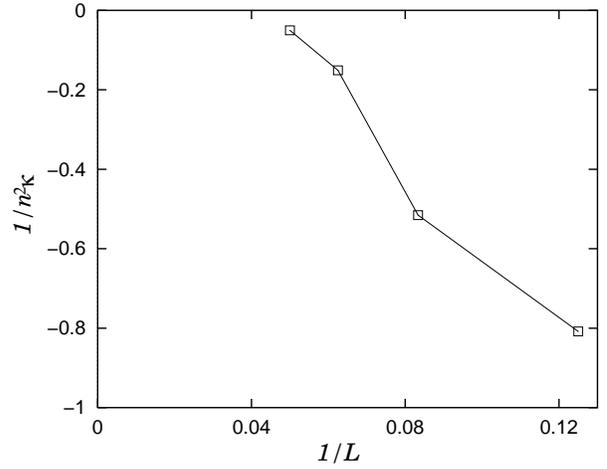}
\caption{Size dependence of the inverse compressibility
in the phase-separated regime at $n=1/2$, $J/t=4$.}\label{fig:ksizep}
\end{figure}
\begin{figure}
\epsfxsize=3.3in \epsfbox{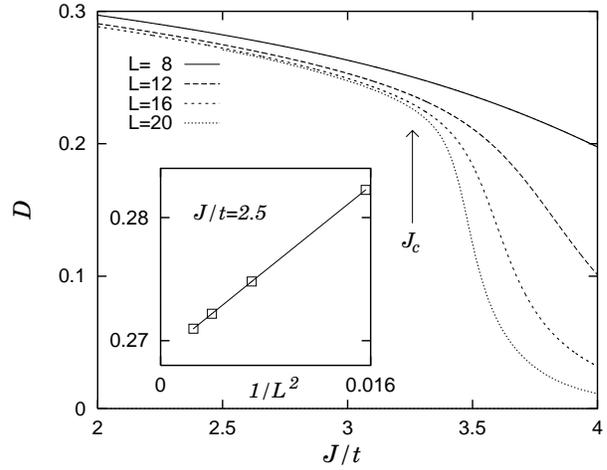}
\caption{Drude weight $D$ as a function of $J/t$
for different system sizes at $n=1/2$.
The arrow indicates the critical point determined
by $\kappa^{-1}=0$.
The size dependence at $J/t=2.5$ is shown in the inset.}\label{fig:Drude1}
\end{figure}
\begin{figure}
\epsfxsize=3.3in \epsfbox{f4z.eps}
\caption{Charge velocity $v_c$ derived by (\protect{\ref{eqn:vc}})
for different system sizes,
the extrapolated one (lines),
and the one derived from (\protect{\ref{eqn:vc2}})
by 16 site data ($\times$),
as a function of $J/t$ at $n=1/2$.}\label{fig:velsc1}
\end{figure}
\begin{figure}
\epsfxsize=3.3in \epsfbox{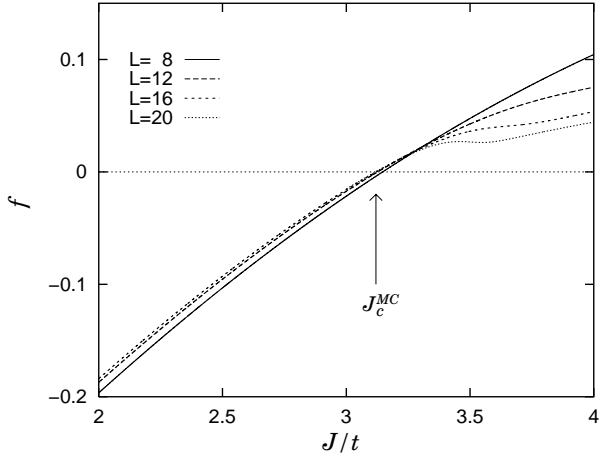}
\caption{Legendre transformed ground state energy density
$f\equiv e -\mu n$ as a function of $J/t$ for different system sizes
at $n=1/2$.
The point $f = 0$ corresponds to the phase-separation boundary
determined by the Maxwell construction.}\label{fig:first}
\end{figure}
\begin{figure}
\epsfxsize=3.3in \epsfbox{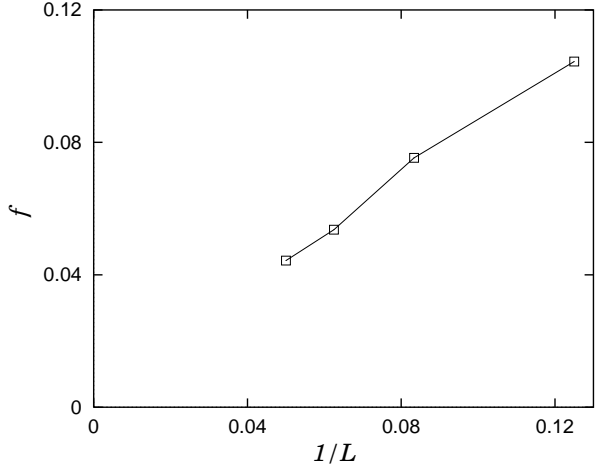}
\caption{Size dependence of the $f\equiv e-\mu n$
in the phase-separated regime at $n=1/2$, $J/t=4$.}\label{fig:fsizep}
\end{figure}
\begin{figure}
\epsfxsize=3.3in \epsfbox{f7z.eps}
\caption{Legendre transformed ground state energy density
$f\equiv e -\mu n$ as a function of the electron density $n$
in $L=16$ system for various values of $J/t$.}\label{fig:grand_ene}
\end{figure}
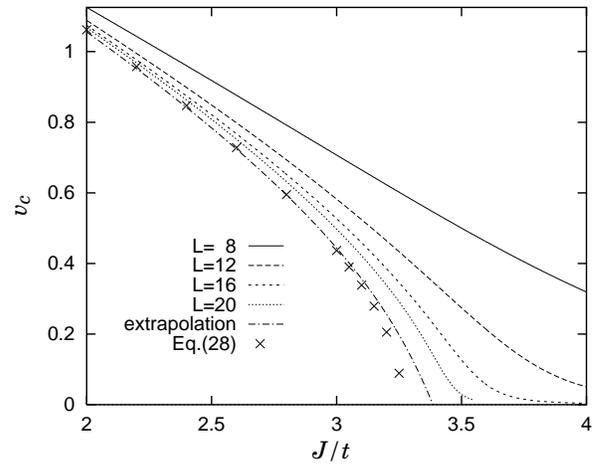
\end{document}